\documentclass[aps,prb,showpacs,groupedaddress,superscriptaddress,twocolumn]{revtex4}

\usepackage{graphicx}
\begin{document}

\title{The spin-polarized $\nu=0$ state of graphene: a spin superconductor}

\author{Qing-feng Sun}
\email{sunqf@iphy.ac.cn}
\affiliation{International Center for Quantum Materials, Peking
University, Beijing 100871, China}
\affiliation{Institute of Physics, Chinese
Academy of Sciences, Beijing 100190, China}

\author{X. C. Xie}
\affiliation{International Center for Quantum Materials and School of Physics, Peking
University, Beijing 100871, China}

\date{\today}

\begin{abstract}
We study the spin-polarized $\nu=0$ Landau-level state of graphene.
Due to the electron-hole attractive interaction, electrons and holes
can bound into pairs. These pairs can then condense into a spin-triplet superfluid
ground state: a spin superconductor state.
In this state, a gap opens up in the edge bands as well as in the
bulk bands, thus it is a charge insulator,
but it can carry the spin current without dissipation.
These results can well explain the insulating behavior of the spin-polarized
$\nu=0$ state in the recent experiments.
\end{abstract}


\maketitle

\section{Introduction}

In a magnetic field, monolayer and bilayer graphenes display
unconventional Landau-level (LL) spectrum, where the zeroth LL
locates the charge neutrality point and has equal electron and
hole compositions.\cite{ref1,ref2,ref3} The zeroth LL is fourfold
degenerate in monolayer graphene owing to the spin and valley
degeneracies, and it is eightfold degenerate in bilayer one due to the
additional orbit (or layer) degeneracy. While under a high
magnetic field, electron-electron (e-e) interaction can lift the LL
degeneracy,\cite{ref2,ref3,ref6,ref7,ref8,ref9,ref10,ref11,ref12,ref13,ref14,ref15}
leading to broken symmetry quantum Hall states and manifesting
further integer Hall plateaus outside the normal sequence, which
have been experimentally
observed.\cite{ref16,ref17,ref18,ref19,ref20,ref21,ref22,ref23,ref24,ref25,ref26,ref27,ref28,ref29,ref30}

Recently, the splitting of the zeroth LL has attracted considerable theoretical and experimental
interest.\cite{ref6,ref7,ref8,ref9,ref10,ref11,ref12,ref13,ref14,ref15,ref16,ref17,ref18,ref19,ref20,ref21,ref22,ref23,ref24,ref25,ref26,ref27,ref28,ref29,ref30,ref31,ref32,ref33,ref34,ref35,ref36}
A bulk gap opening around the energy $E=0$ is found and a zero Hall conductance plateau at the filling factor $\nu=0$ has been observed.
Both the spin-polarized and valley-polarized $\nu=0$ states are suggested.
At $\nu=0$, although the Hall conductance shows a
plateau, the longitudinal resistance experimentally exhibits an
insulating behavior,\cite{ref16,ref17,ref18,ref19,ref20,ref21,ref22,ref23,ref24,ref25,ref26,ref27,ref28,ref29,ref30}
which is very different with the zero longitudinal resistance in the conventional quantum Hall effect.

In the valley-polarized $\nu=0$ state, the valley splitting is larger than the spin splitting
and it is a spin singlet state. Now not only $\nu=0$, but also the
spin-up and spin-down filling factors
$\nu_{\uparrow}=\nu_{\downarrow}=0$.\cite{ref36} In this case, the
system is without an edge state as shown in Fig.1(a), so it is
insulating for both bulk and edge states, which is consistent with the experiment results.

On the other hand, when the
spin splitting is larger than the valley splitting, the system is in the spin-polarized $\nu=0$
state.\cite{ref7,ref31}
Now, however, $\nu_{\uparrow}$ and
$\nu_{\downarrow}$ are not equal to zero although
$\nu=\nu_{\uparrow}+\nu_{\downarrow}=0$. A $+$ valley spin-up
($+\uparrow$) LL is occupied by electron and a $-$ valley spin-down
($-\downarrow$) LL is occupied by hole, leading to a pair of
counter-propagating edge states [see Fig.1(b)] that can carry both
spin and charge
currents.\cite{ref31,ref32,ref33,ref34,ref35,ref36} Some theoretical
works have predicted the spin Hall effect in this
case.\cite{ref34,ref35} Particularly, due to the presence of the edge
states, the longitudinal resistance is
$(|\nu_{\uparrow}|+|\nu_{\downarrow}|)e^2/h$ and the system should not
show an insulating behavior, although a bulk gap exists. However,
experiment works have clearly exhibited an insulating behavior and the
longitudinal resistance increases quickly with decreasing temperature
regardless whether it is a monolayer or bilayer
graphene.\cite{ref17,ref20,ref21,ref27,ref28,ref30} This is
very different from the theoretical prediction and seems to indicate the disappearance of
the edge states.
Some studies mention that the possible reason for this discrepancy is
that the counter-propagating edge states are destroyed by
disorders.\cite{ref27,ref32,ref33,ref35,ref36} But the sizes of
the experimental samples are only a few micrometers, too short to destroy
the edge states by localizing the edge electrons. Furthermore, the disorder effect can not explain strong
increase of the longitudinal resistance at low temperature. In
short, this discrepancy is still lack of a reasonable explanation.

In this paper, the spin-polarized $\nu=0$ state in graphene under a
strong magnetic field is investigated. By considering the unavoidable electron-hole
(e-h) attractive interaction, we find that electrons at $+\uparrow$
LL and holes at $-\downarrow$ LL can form spin-triplet e-h pairs.
This e-h pair gas can condense at low
temperature, leading to the transition to a spin
superconductor phase (spin-triplet exciton condensation state)\cite{ref37} associated with the opening of an energy gap for the edge states.
Thus, the system exhibits an insulating behavior, consistent
with experimental
observations.\cite{ref17,ref20,ref21,ref27,ref28,ref30}

The remainder of this paper is organized as follows.
In Sec. II,
we introduce the Hamiltonian in the tight-binding representation
and derive the formula of the spin-superconductor order parameter.
The results are discussed in Sec. III.
Finally, the conclusion is presented in Sec. IV.

\section{Model and formulation}

Let us consider a graphene nanoribbon in a magnetic field. In the
tight-binding representation, its Hamiltonian is
$\mathcal{H}=\mathcal{H}_0+\mathcal{H}_I$, where
\begin{eqnarray}
\mathcal{H}_0 &=& \sum\limits_{{\bf i}, \sigma} (\epsilon_{\bf i}
 -\sigma M) a^{\dagger}_{{\bf i}\sigma} a_{{\bf i}\sigma}
 +\sum\limits_{{\bf i},{\bf i}', \sigma} t_{{\bf i}{\bf i}'}
 e^{i\phi_{{\bf i}{\bf i}'}}
 a^{\dagger}_{{\bf i}\sigma} a_{{\bf i}'\sigma} +H.c.  \nonumber\\
\mathcal{H}_I &=& \sum\limits_{{\bf i},{\bf i}',\sigma,\sigma' ({\bf i}\sigma\not={\bf i}'\sigma')}
 U_{{\bf i}{\bf i}'}
 a^{\dagger}_{{\bf i}\sigma} a_{{\bf i}\sigma}
 a^{\dagger}_{{\bf i}'\sigma'} a_{{\bf i}'\sigma'} ,
\end{eqnarray}
represent the free part and the e-e Coulomb
interaction part of the Hamiltonians, respectively. Here $a^{\dagger}_{{\bf i}\sigma}$
($a_{{\bf i}\sigma}$) is the electron creation (annihilation)
operator at sites ${\bf i}$ with spin $\sigma$. $\epsilon_{\bf i}$ is
the on-site energy, and $M$ is spin splitting energy which origins from
both the Zeeman effect and the spin
polarization induced by the e-e interaction. The second term in $\mathcal{H}_0$
represents the hopping between the site ${\bf i}$ and ${\bf i}'$.
Because of the presence of a magnetic field $B$, a phase $\phi_{{\bf i}{\bf i}'} =\int_{{\bf i}}^{{\bf i}'} {\bf A} \bullet d{\bf l} /\phi_0$ ($\phi_0=\hbar/e$) is
attached in the hopping element $t_{{\bf i}{\bf i}'}$.\cite{ref38}
$\mathcal{H}_I$ is the e-e interaction and $ U_{{\bf
i}{\bf i}'}$ is the interaction strength.
This Hamiltonian $\mathcal{H}$ can describe both monolayer and
bilayer graphene ribbons with arbitrary edge chirality. Considering
the ribbon periodicity, the site indices ${\bf i}$ can be
represented as ${\bf i}=(n,j)$ with the slice cell indices $n$ and
the atomic indices $j$ in a cell ($j=1,2,\ldots, N$ and $N$ is the
total atom number in a cell). Then the Hamiltonian can be
rewritten as:
\begin{eqnarray}
 \mathcal{H}_0 &=& \sum\limits_{n,\sigma}
 \vec{a}^{\dagger}_{n\sigma} ({\bf H}_0 -\sigma M) \vec{a}_{n\sigma}
 + \sum\limits_{n,\sigma}
 \vec{a}^{\dagger}_{n\sigma} {\bf H}_{1} \vec{a}_{n-1 \sigma}  +H.c. \nonumber\\
 \mathcal{H}_I &=& \sum\limits_{n,n',\sigma,\sigma'}
 \vec{n}^{\dagger}_{n\sigma} {\bf U}_{n-n'}
 \vec{n}_{n'\sigma'},
\end{eqnarray}
where $\vec{a}_{n\sigma}=(a_{n1}, a_{n2}, \ldots, a_{nN})^T$ and
$\vec{n}_{n\sigma}=(a^{\dagger}_{n1}a_{n1}, a^{\dagger}_{n2}a_{n2},
\ldots, a^{\dagger}_{nN}a_{nN})^T$. ${\bf H}_0$, ${\bf H}_{1}$, and
${\bf U}_{n-n'}$ are the intra-cell Hamiltonian,
hopping term between two nearest-neighbor cells, and the e-e
interaction. By taking the Fourier transformation $\vec{a}_{n\sigma}
= \frac{1}{\sqrt{L}} \sum_k e^{inak} \vec{a}_{k\sigma}$ with the
nanoribbon length $L a$ and the cell length $a$, the Hamiltonian
$\mathcal{H}$ can be written as: $\mathcal{H}_0= \sum_{k,\sigma}
\vec{a}^{\dagger}_{k\sigma}({\bf H}_{k} -\sigma M)
\vec{a}_{k\sigma}$ with ${\bf H}_{k}= {\bf H}_0 +{\bf H}^{\dagger}_1
e^{iak} +{\bf H}_1 e^{-iak}$, and
\begin{equation}
 \mathcal{H}_I = \sum\limits_{\sigma,\sigma', k, k',q} \vec{a}^{\dagger}_{k-q \sigma}
 \vec{a}_{k \sigma} {\bf U}_{k-k'}
 \vec{a}^{\dagger}_{k'+q \sigma'}
 \vec{a}_{k' \sigma}
\end{equation}
with $ {\bf U}_{k} =\frac{1}{L} \sum_n e^{inak} {\bf
U}_{n}$. In fact, ${\bf H}_{k}$ is the momentum-space
Hamiltonian of the free system. Assuming that the eigen-wavefunctions and eigenvalues of ${\bf
H}_k$ are $\vec{\Psi}^{(j)}_k$ and $\epsilon^{(j)}_k$: ${\bf H}_k
\vec{\Psi}^{(j)}_k = \epsilon^{(j)}_k\vec{\Psi}^{(j)}_k$, we have
${\bf \mathcal{U}}_k^{\dagger} {\bf H}_{k} {\bf \mathcal{U}}_k ={\bf
\epsilon}_k$ with ${\bf \mathcal{U}}_k
=(\vec{\Psi}^{(1)}_k,\vec{\Psi}^{(2)}_k,\ldots,\vec{\Psi}^{(N)}_k)$
and ${\bf \epsilon}_k
=diag(\epsilon^{(1)}_k,\epsilon^{(2)}_k,\ldots, \epsilon^{(N)}_k)$.
By taking a unitary transformation: $\vec{a}_{k\sigma}= {\bf
\mathcal{U}}_k \vec{b}_{k\sigma}$, $\mathcal{H}_0$ changes into
$\mathcal{H}_0 =\sum_{k\sigma} \vec{b}^{\dagger}_{k\sigma} ( {\bf
\epsilon}_k -\sigma M ) \vec{b}_{k\sigma}$.

Let us assume that the eigenvalues $\epsilon^{(j)}_k$ have been
arranged according of their values from small to large and Fermi
level $E_F$ is set at zero. The two nearest bands to $E_F$ are the
spin-up $1+N/2$-th band and spin-down $N/2$-th band. Due to the
presence of a magnetic field and the spin splitting energy $M$, the system consists of LLs
and is spin-polarized. The spin-up $1+N/2$-th (spin-down $N/2$-th)
band is denoted as $+\uparrow$ ($-\downarrow$) LL with its energy
$\epsilon^{(1+N/2)}_k -M$ below ($\epsilon^{(N/2)}_k +M$ above)
$E_F$, its carrier being electron-like (hole-like), and its band bending
upward (downward) as shown in Fig.1(b). Now the system is at the
spin-polarized $\nu=0$ state, in which a bulk gap $2M$ appears but two
edge states cross at $E_F$. In the
following, we focus on these two low-energy bands, and show
that the e-e interaction $\mathcal{H}_I$ will create an energy gap for the
edge states. Let us introduce the electron and hole annihilation
operators: $b_{k\uparrow e} =b_{k\uparrow, 1+N/2}$ and $b_{k\uparrow
h} =b^{\dagger}_{k\downarrow, N/2}$. Then the free Hamiltonian
$\mathcal{H}_0$ reduces to:
$$
 \mathcal{H}_0 = \sum\limits_{k}
 [ b^{\dagger}_{k\uparrow e} (\epsilon_{ke} -M)
 b_{k\uparrow e} + b^{\dagger}_{k\uparrow h} (\epsilon_{kh} - M)
 b_{k\uparrow h} ]
$$
with $\epsilon_{ke} = \epsilon^{(1+N/2)}_{k}$ and $\epsilon_{kh} = -\epsilon^{(N/2)}_{k}$.

As for $\mathcal{H}_I$, we take the
following steps: 1) only the terms whose momenta satisfy
$k=k'+q$ in Eq.(3) are kept since the zero momentum e-h pairs
are energetically more favorable; 2) we take the aforementioned unitary
transformation and the e-h transformation; 3) we focus on the two
low-energy bands; and 4) we assume that $U_{{\bf i}{\bf i}'} =U_c$
while ${\bf i}={\bf i}'$ and $U_{{\bf i}{\bf i}'} =0$ otherwise,
since the on-site e-e interaction is the dominant one. Then the
interaction part $\mathcal{H}_I$ reduces to:
\begin{equation}
 \mathcal{H}_I = - \sum\limits_{k, k'}
  U_{kk'} b^{\dagger}_{k'\uparrow e} b^{\dagger}_{k'\uparrow h}
 b_{k \uparrow h} b_{k\uparrow e},
\end{equation}
where
$$U_{kk'}= \frac{1}{L} \sum_j \mathcal{U}_{k',jN/2}^*
\mathcal{U}_{k,jN/2} U_c \mathcal{U}_{k,j 1+N/2}^* \mathcal{U}_{k',j
1+N/2} $$.

While at equilibrium, the spin-up electrons (holes) occupy the $+\uparrow$
($-\downarrow$) LL and its edge state up to the energy $E=E_F=0$.
This is a spin-polarized $\nu=0$ state and has $U(1)$ symmetry around
$\sigma_z$-axis. Notice that the interaction $\mathcal{H}_I$ in
Eq.(4) between an electron and a hole is attractive.
This attractive
interaction will not cause the e-h recombination, due to both the spin splitting and the
hole band ($-\downarrow$ LL) is above the electron band ($+\uparrow$
LL).\cite{ref37}
However, it can lead to a different instability of the
spin-polarized $\nu=0$ state at low temperature, namely the electrons
and holes can form e-h pairs which can then condense to a spin-triplet superfluid
state.\cite{ref37,ref39}
Notice here the spin splitting (or spin polarization) is a key factor for stable e-h pairs.
Under the mean-field approximation, $\mathcal{H}_I$ changes into:
\begin{equation}
 \mathcal{H}_I = \sum\limits_{k}
 \left[\Delta_k  b^{\dagger}_{k\uparrow e} b^{\dagger}_{k\uparrow h}
 + \Delta_k^*  b_{k\uparrow h} b_{k\uparrow e}\right],
\end{equation}
where $\Delta_k \equiv -\sum_{k'} U_{kk'} \langle b_{k'\uparrow h} b_{k'
\uparrow e} \rangle$ is the e-h pair condensation order parameter.
So we have the total Hamiltonian $\mathcal{H}= \mathcal{H}_0
+\mathcal{H}_I$:
\begin{equation}
 \mathcal{H} = \sum\limits_{k}
  \left( \begin{array}{l}
  b^{\dagger}_{k\uparrow e},  b_{k\uparrow
  h}\end{array}\right)
  \left( \begin{array}{ll}
   \epsilon_{ke}-M & \Delta_k \\
   \Delta_k^* & M-\epsilon_{kh}  \end{array} \right)
  \left(\begin{array}{l}
    b_{k\uparrow e} \\ b^{\dagger}_{k\uparrow  h}\end{array}\right).
\end{equation}

Now a gap $|\Delta_k|$ opens up in the edge bands [e.g. see Fig.1(c)
and (d)], and it needs an energy $2|\Delta_k|$ to break up an e-h
pair. So the e-h pair condensed state is stabler than the
spin-polarized $\nu=0$ state and it is the ground state of the
system at low temperature. Since the spins of the electrons and
the holes are both up, the e-h pair is spin triplet but charge neutral.
The condensed superfluid state is a spin superconductor
while it is a charge insulator.\cite{ref37,ref40} It carrys spin current
dissipationlessly, thus, its spin resistance is zero.
The spin superconductor also posseses its own unique
'Meissner effect".\cite{ref37}
Now the system has two possible phases. One is the spin-polarized $\nu=0$
state (hereafter we named it as normal state for short) at high temperature. It has a bulk
gap but two gapless edge bands crossover at the Fermi level, leading
to the current flow through the edge
states.\cite{ref31,ref32,ref33,ref34,ref35,ref36} In the normal
phase, the system consists of $U(1)$ symmetry. The other is the spin superconductor
state at low temperature, in which both bulk bands and edge
bands consist of energy gaps at $E_F$. Notice that this phase is still a spin polarized
one and its filling factor $\nu=0$ with
$\nu_{\uparrow}=-\nu_{\downarrow}=1$. We name it as spin-superconductor
spin-polarized $\nu=0$ state, or spin superconductor state for
short. This phase does not contain $U(1)$ symmetry around any
direction. In other words, the system breaks $U(1)$ symmetry with
the phase transition from the normal phase to the spin superconductor phase.

\section{Results and discussions}

From the definition of $\Delta_k$ and Hamiltonian (6), we obtain
the the self-consistent equation of $\Delta_k$:
$
 \Delta_k = -i \sum_{k'} U_{k'k} \int \frac{d\epsilon}{2\pi}
 f(\epsilon) ( \frac{\Delta_{k'}}{A} -\frac{\Delta_{k'}}{A^*} ),
$
where $f(\epsilon)$ is Fermi distribution function and
$A=(\epsilon-\epsilon_{k'e}+M+i0^+)(\epsilon +\epsilon_{k' h}-M
+i0^+)-|\Delta_{k'}|^2$. While at zero temperature, the above
equation reduces to:
\begin{equation}
 \Delta_k = \sum\limits_{k'} U_{k'k} \frac{\Delta_{k'}}{\sqrt{(\epsilon_{k'e}
 +\epsilon_{k'h} -2M)^2 +4 |\Delta_{k'}|^2}}.
\end{equation}

From this equation, $\Delta_k$ can be self-consistently calculated.
In the numerical calculations, we first consider the monolayer zigzag
graphene ribbon with the ribbon transverse width $W=(3N/4-1)a_0$ and
periodic cell length $a=\sqrt{3}a_0$. Here $a_0=0.142$nm is the distance
between two nearest-neighbor carbon atoms. We only consider the
nearest-neighbor hopping with its strength $t_{{\bf i}{\bf
i}'}=t=2.75$eV, which is set as the energy unit. The on-site e-e
interaction $U_c =\frac{e^2}{4\pi\epsilon_0 r} \approx
\frac{3.69}{r/a_0} t$ with the distance $r$ between two electrons.
$U_c\approx 14.75 t$ if $r=a_0/4$.

\begin{figure}
\includegraphics[width=8.5cm,totalheight=12cm]{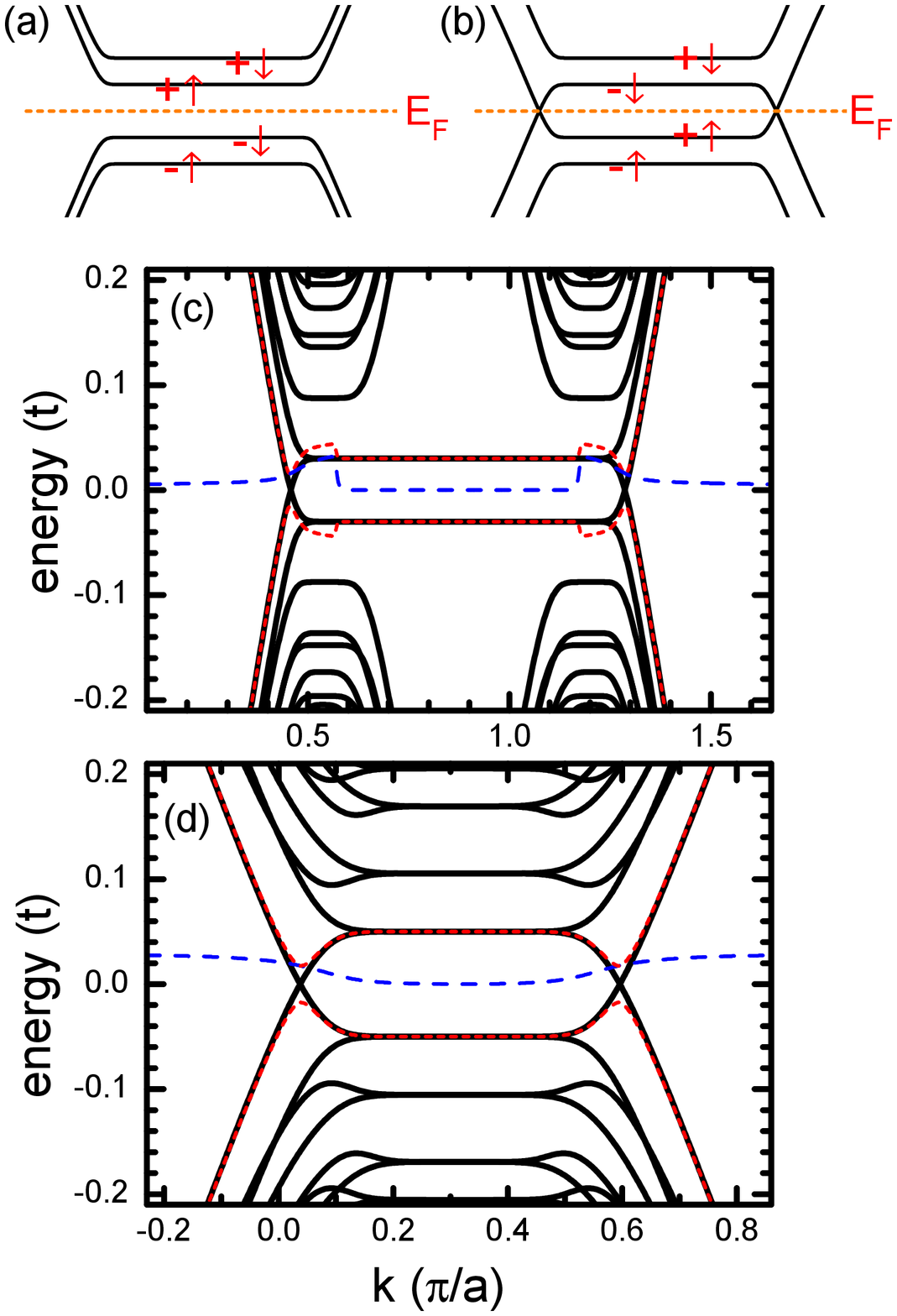}
\caption{ (color online) (a) and (b) are the schematic diagrams of
the energy spectrum structures for the valley-polarized and
spin-polarized $\nu=0$ states, respectively. (c) and (d) show
$\Delta_k$ (the blue dashed curves), and the energy spectra
for the normal state (the black solid curves) and for the
spin superconductor state (the red dotted curves). (c) is for the
zigzag edge graphene ribbon with the parameters $N=200$, $U_c=15t$,
$M=0.03t$, and $\phi=0.004$, and (d) is for the armchair edge
graphene ribbon with the parameters $N=282$, $U_c=25t$, $M=0.05t$,
and $\phi=0.007$. }
\end{figure}

Fig.1(c) shows $\Delta_k$ and the energy spectrum. For the normal
state, although it has a bulk gap due to the spin splitting energy
$M$, two gapless edge states cross at the Fermi level and they can
carry both charge and spin currents, causing
the sample edge having a metallic
behavior.\cite{ref31,ref32,ref33,ref34,ref35,ref36} On the other
hand, for the spin superconductor state at low temperature,
Fig.1(c) clearly exhibits a gap opening for the edge bands. Now
both edge and bulk bands have the gaps, so it is a charge
insulator, consistent with the experimental
results.\cite{ref16,ref17,ref18,ref19,ref20,ref21,ref22,ref23,ref24,ref25,ref26,ref27,ref28,ref29,ref30}
In this state, the spin current can dissipationlessly flow in it,
because the condensed e-h pairs with spin $1$ can carry
the spin super-current. Except for the edge states, other parts of the
bands are almost the same for both normal and spin superconductor
states and their LLs overlap, because that the carriers far away
from Fermi level are not energetically favorable to form the e-h
pairs. $\Delta_k$ is large for the edge bands but is vanishingly small for
the bulk bands. This means that the condensed e-p pairs mainly
distribute near the sample edge, and the spin super-current flows
along the edges.

Up to now, we only consider the monolayer zigzag edge graphene. In
fact, it is similar for graphene with other edge chirality as well as for
a bilayer graphene.
For example,
Fig.1(d) shows $\Delta_k$ and the energy spectrum for the
armchair edge graphene nanoribbon with the ribbon width
$W=\frac{N-2}{4}\sqrt{3}a_0$ and cell length $a=3a_0$. The free
Hamiltonian exhibits two gapless edge states.
The e-h attractive interaction induces a gap in the edge
states at the low temperature. The order
parameter $\Delta_k$ is large for the edge bands but is very small for
the bulk bands.

\begin{figure}
\includegraphics[width=8.5cm,totalheight=6cm]{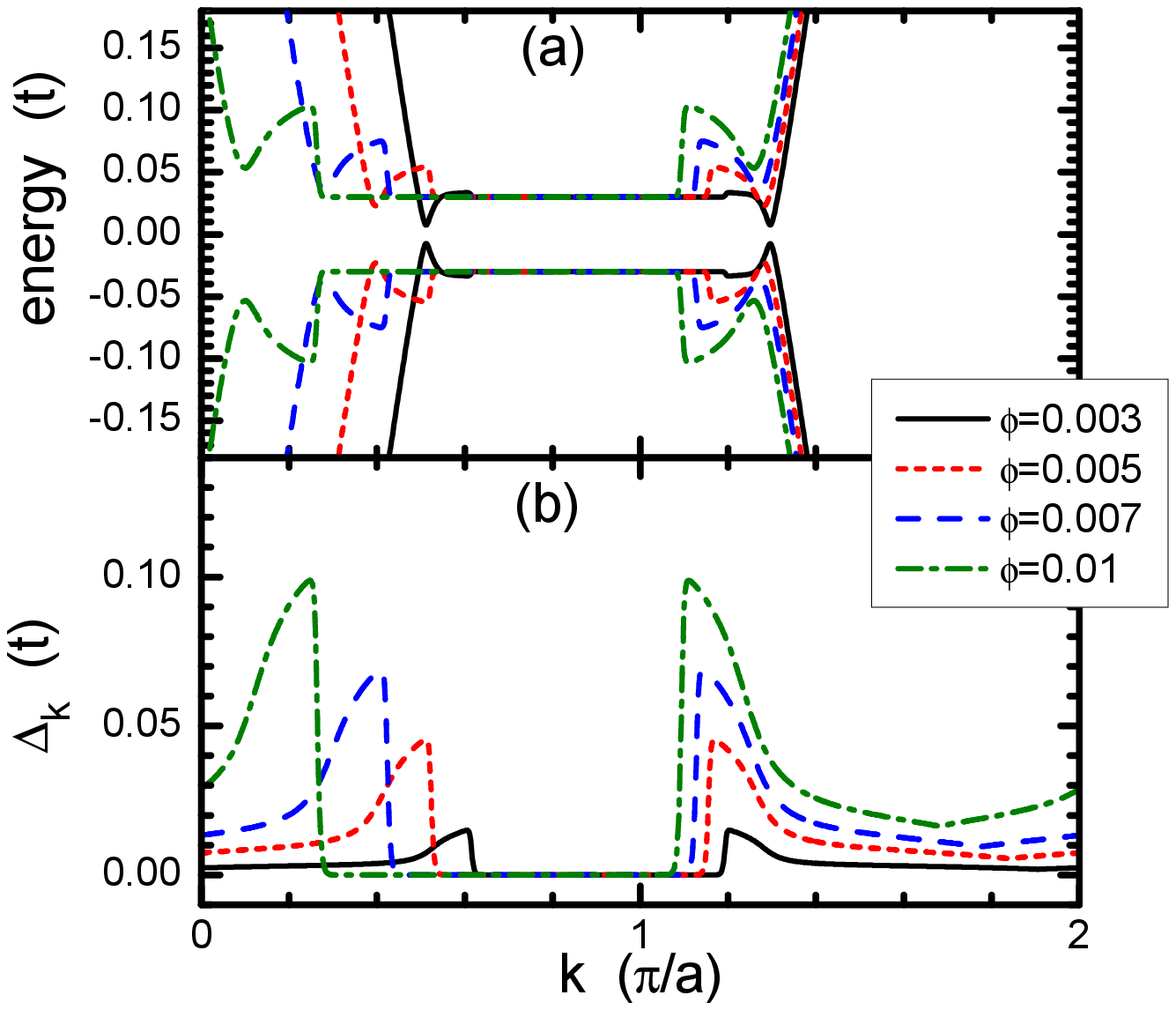}
\caption{ (color online) (a) and (b) show the energy spectra
of the spin superconductor and the order parameter
$\Delta_k$, respectively. The parameters are the same as in Fig.1(c). }
\end{figure}

Next, we study the zigzag edge
graphene nanoribbon in detail. Fig.2(a) and 2(b) show the energy
spectrum of spin superconductor state and $\Delta_k$ for
different magnetic fields $\phi$ (here $2\phi =(3\sqrt{3}/2)a_0^2 B/\phi_0$ is the magnetic flux
in the honeycomb lattice). For all $\phi$, $\Delta_k$ exhibits
peaks when the original bands cross at $E_F$ and
$\Delta_k$ is small otherwise. With increasing $\phi$,
$\Delta_k$ increases because
a larger magnetic field leads to a smaller cyclotron radius of carriers,
thus a stronger e-h attractive interaction
$U_{k'k}$. Particularly, for a large $\phi$, the edge-band gap can
exceed the bulk-band gap (i.e. $2M$).
In this case, the edge
states disappear in the whole spin-polarized $\nu=0$ regime, as has
been observed in the
experiments.\cite{ref17,ref20,ref21,ref27,ref28,ref30}

Fig.3 shows the energy spectrum of spin superconductor state for
different width of nanoribbon. The results exhibit that both the
edge-band gap and bulk-gap gap are almost independent with the
width $N$. Because while under the high magnetic field the edge
states and LLs in the spin-polarized state are independent
with the width of nanoribbons.

\begin{figure}
\includegraphics[width=8.5cm,totalheight=5.5cm]{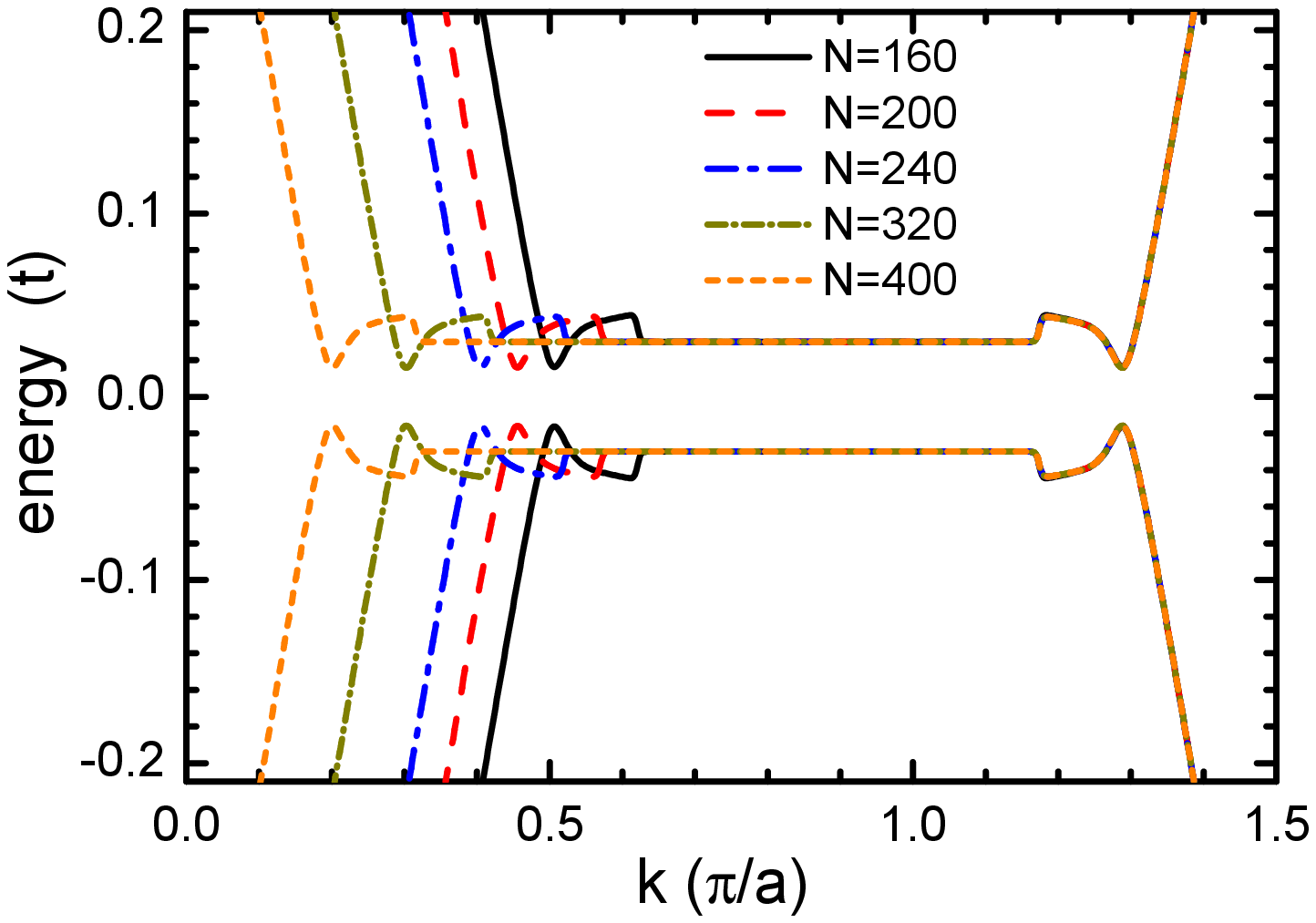}
\caption{ (color online) The energy spectra for the spin
superconductor state for the different width of nanoribbon.
The parameters are the same as in Fig.1(c). }
\end{figure}

Let us study the effect of the system parameters on the energy gap. Here
the gap is defined as the one of whole energy spectrum, equal to
the smaller one of the bulk and the edge gaps.
The gap is independent of the width $N$ of nanoribbon.
Fig.4(a), (b), and (c) show
the gap versus the spin splitting energy $M$, the e-e interaction strength
$U_c$, and the magnetic flux $\phi$, respectively. With increase of
$M$, the gap first increases due to the rising of the bulk gap, and
then decreases. As for the gap versus $U_c$, there exists a
threshold $U_c^t$ [see Fig.4(b)]. While
$U_c<U_c^t$, the gap is almost zero,
but while $U_c>U_c^t$, the gap increases quickly. When the gap
reaches the bulk gap $2M$, it hardly increases further. In this case,
although the edge gap can further increase, the gap of whole energy
spectrum is decided by the bulk gap. The results of the gap vs.
$\phi$ is similar to the one of the gap vs. $U_c$ since
the increase of $\phi$ strengthens the effective e-h interaction
$U_{k k'}$.
In an experiment, normally the spin splitting
energy $M$ linearly rises with a magnetic field
$B$. So in Fig.4(d), we show the gap vs. $\phi$ while $M=2\mu_B B$
and $4\mu_B B$. The results clearly exhibit that the gap linearly
rises and the edge gap is always larger than the bulk gap. Now the
edge states disappear in all $\phi$ value, which is well consistent
with the experiment
results.\cite{ref16,ref17,ref18,ref19,ref20,ref21,ref22,ref23,ref24,ref25,ref26,ref27,ref28,ref29,ref30}
While $\phi=0.001$, $B$ is about 25 Tesla, and the gaps are about 3meV and 6meV for $M=2\mu_B B$
and $4\mu_B B$ respectively, to give rise to the corresponding critical temperatures of the phase transition to be about 30K and 60K.

\begin{figure}
\includegraphics[width=8.5cm,totalheight=6cm]{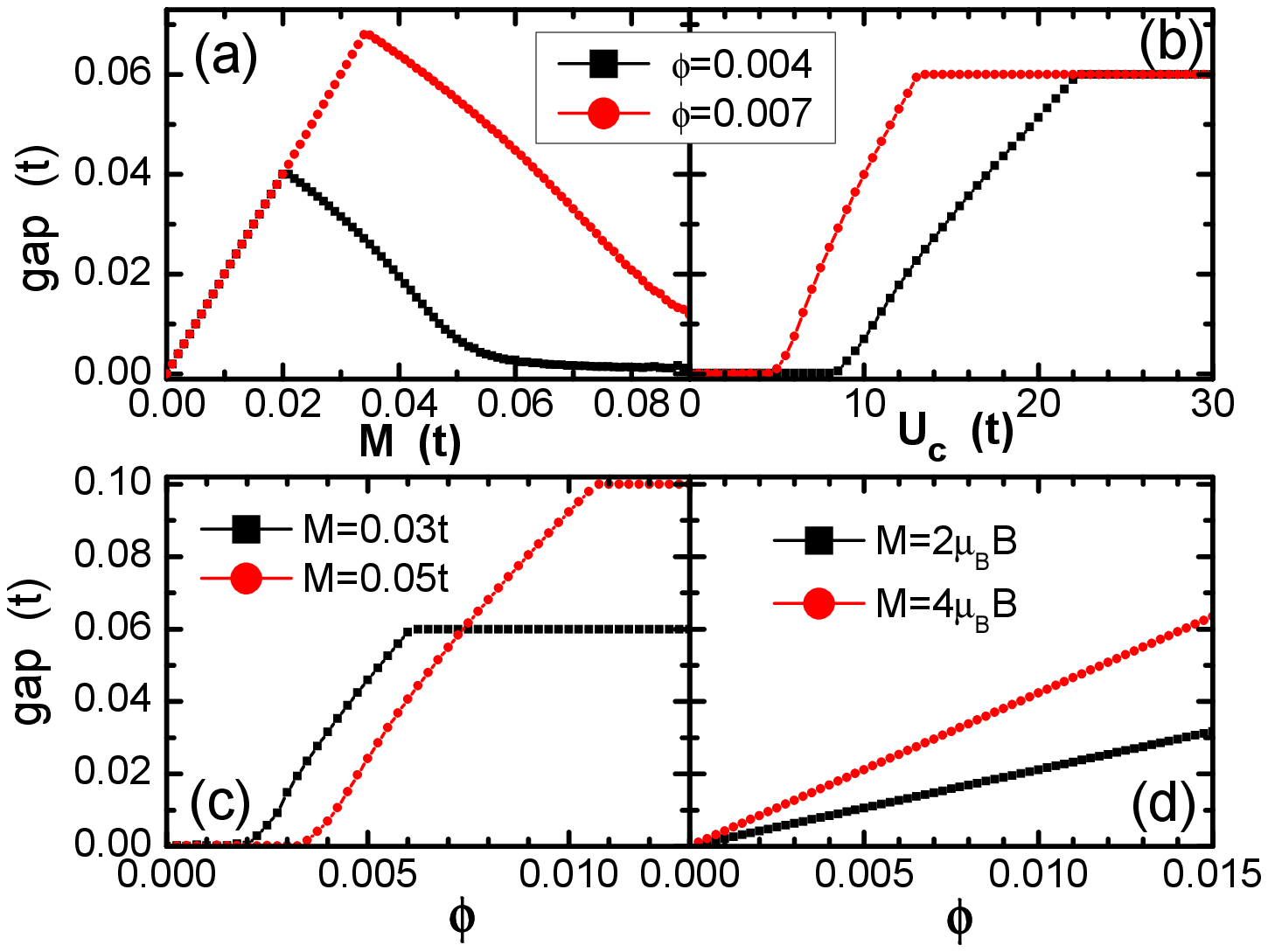}
\caption{ (color online) The gap versus $M$
(a), $U_c$ (b), and $\phi$ (c and d).
The unmentioned parameters are the same as Fig.1(c). }
\end{figure}

Finally, we notice that a recent experiment has simultaneously measured the
resistance and the nonlocal resistance in graphene under a magnetic
field.\cite{ref21} They find that the device is insulator at
neutrality point with $\nu=0$.
But, they also see that the nonlocal resistance
increases rapidly at low temperature,
and shows clearly that a spin
current is flowing through the device.
These findings can well be explained
by the presence of a spin superconducting $\nu=0$ state.

\section{conclusions}

In summary, the spin-polarized $\nu=0$ state of the graphene under a
magnetic field is investigated. We find that it has two phases, one
is the normal phase at high temperature
and other is the spin superconductor phase at
low temperature. The $U(1)$ symmetry is destroyed under the phase transition
from the normal phase to a spin superconductor. For the spin superconductor phase, both edge and bulk
bands contain gaps, so it is a charge insulator, but the spin
current can flow without dissipation. With the picture of
the spin superconductor, many results from
recent experiments can be well understood.

\section*{Acknowledgments}
This work was financially supported by NBRP
of China (2012CB921303, 2009CB929100 and 2012CB821402), NSF-China, under Grants
Nos. 11074174, 11121063, 91221302 and 11274364.

\end{document}